\documentclass[aps,prl,twocolumn,showpacs]{revtex4}

\usepackage{amsmath,amssymb}
\usepackage{graphicx}
\usepackage{fancybox,color}

\usepackage[latin1]{inputenc} % Umalaute

\def\hc{\text{h.c.}}
\def\d{\partial}
\def\h{\hat}

\def\s{{}^\dagger}

\newcommand{\nn}{\nonumber\\}

\newcommand{\bea}{\begin{eqnarray}}
\newcommand{\ea}{\end{eqnarray}}
\newcommand{\eea}{\end{eqnarray}}

\begin{document}
%%%%%%%%%%%%%%%%%%%%%%%%%%%%%%%%%%%%%%%%%%%%%%%%%%%%%%%%%%%%%%%%%%%%%%%%%%%%%%%

\title{Quantum simulator for the Schwinger effect with 
atoms in bi-chromatic optical lattices}

\author{Nikodem Szpak}
\email[e-mail:]{nikodem.szpak@uni-due.de}
\author{Ralf Sch{\"u}tzhold}
\email[e-mail:]{ralf.schuetzhold@uni-due.de}

\affiliation{Fakult{\"a}t f{\"u}r Physik, Universit{\"a}t Duisburg-Essen,
Duisburg, Germany}

\date{\today}

\begin{abstract}
Ultra-cold atoms in specifically designed optical lattices can be used to 
mimic the many-particle Hamiltonian describing electrons and positrons 
in an external electric field.
This facilitates the experimental simulation of (so far unobserved)
fundamental quantum phenomena such as the Schwinger effect, i.e.,
spontaneous electron-positron pair creation out of the vacuum by 
a strong electric field.
\end{abstract}

\pacs{
67.85.-d, % Ultracold gases, trapped gases
03.65.Pm, % Relativistic wave equations (Quantum mechanics) 
12.20.-m. % Quantum electrodynamics
}

%%%%%%%%%%%%%%%%%%%%%%%%%%%%%%%%%%%%%%%%%%%%%%%%%%%%%%%%%%%%%%%%%%%%%%%%%%%%%%%
\maketitle
%%%%%%%%%%%%%%%%%%%%%%%%%%%%%%%%%%%%%%%%%%%%%%%%%%%%%%%%%%%%%%%%%%%%%%%%%%%%%%%

\paragraph{Introduction}
There are several fundamental predictions of relativistic quantum field 
theory which have so far eluded a direct experimental verification.
One prominent example is the Schwinger \cite{Schwinger} effect
(historically more accurate would be the name Sauter-Schwinger effect,
see \cite{Sauter} and \cite{Heisenberg+Euler}),
i.e., the spontaneous creation of electron/positron-pairs out
of the vacuum by a strong electric field.
For a constant electric field $E$, the leading-order $e^+e^-$ pair creation
probability scales as \cite{Sauter,Heisenberg+Euler,Schwinger}
\bea
\label{probability}
P_{e^+e^-}
\sim
\exp\left\{-\pi\,\frac{c^3}{\hbar}\,\frac{M^2}{qE}\right\}
=
\exp\left\{-\pi\,\frac{E_{\text{S}}}{E}\right\}
\,,
\ea
where ${E_{\text{S}}}=M^2c^3/(\hbar q)$ is the critical field strength
determined by the elementary charge $q$ and the mass $M$ of an electron 
(or positron).
The above expression (\ref{probability}) for $P_{e^+e^-}$ does not permit
a Taylor expansion in $q$, i.e., it is inherently non-perturbative and
thus cannot be represented by any finite set of Feynman diagrams.

Unfortunately, our theoretical understanding of this non-perturbative QED
effect is still very incomplete.
Apart from the constant field case, only very simple field configurations
where the electric field either depends on time $E(t)$ or on one spatial
coordinate such as $E(x)$ are fully solved.
For example, recently it has been found that the occurrence of two
different frequency scales in a time-dependent field $E(t)$ can induce
drastic changes in the (momentum dependent) pair creation probability
\cite{catalysis,Stokes}.
Moreover, the impact of interactions between the electron and the
positron of the created pair, as well as between them and other
electrons/positrons is not understood.
This ignorance is unsatisfactory not only from a theory point of view
but also in view of planned experiments which envisage field strengths 
not too far below the critical field strength ${E_{\text{S}}}$ and thus 
could be able to probe this effect experimentally \cite{ELI}. 

These considerations motivate the investigation of the Schwinger
effect via a different line of approach.
By suitably designing a laboratory system, we could reproduce the
quantum many-particle Hamiltonian describing electrons and positrons 
in an electric field and thereby obtain a quantum simulator for the
Schwinger effect.
This would facilitate the investigation of space-time dependent
electric fields such as $E(t,x)$ and should also provide some insight 
into the role of interactions.
It should be stressed here that our proposal goes beyond the simulation
of the (classical or first-quantized) Dirac equation on the single-particle 
level, see, e.g., 
\cite{Witthaut+Weitz-DoublePeriodicOptPot, RS+Unruh-SlowLight, Dirac-photons, 
Dirac+Zitterbewegung-photons, QSim-Dirac, QSim-Dirac-TrappedIon, 
Schaetz-QSim-Dirac-TrappedIon},
but aims at the full quantum many-particle Hamiltonian. 
A correct description of many-body effects such as particle-hole creation 
(including the impact of interactions) requires creation and annihilation 
operators in second quantization.
There are some proposals for the second-quantized Dirac Hamiltonian 
\cite{QSim-Dirac-HexLattice, Dirac+Interaction-OptLat,
MasslessDirac-SqLattice, Dirac-StaggeredMagnField,
Goldman+Lewenstein-Dirac-by-SU2, Lewenstein-Dirac-CurvedST}
but they consider scenarios which are more involved than the set-up 
discussed here and aim at different models and effects. 
Similarly, the recent observation of Klein tunnelling in graphene
\cite{Novoselov+Geim-Graphene-Dirac, Novoselov+Geim-Graphene-KleinPar} 
deals with massless Dirac particles -- but the mass gap is crucial for the 
non-perturbative Schwinger effect, cf.~Eq.~(\ref{probability}). 

\paragraph{The model} 
We start with the Dirac equation \cite{Dirac} describing electrons and 
positrons propagating in an electromagnetic vector potential 
$A_\mu$ which are described by the spinor wave-function $\Psi$
($\hbar=c=1$)
\begin{equation}
\gamma^\mu (i\d_\mu-qA_\mu) \Psi - M\, \Psi = 0
\,.
\end{equation}
For simplicity, we consider 1+1 dimensions ($\mu=0,1$) where the
Dirac matrices $\gamma^\mu$ satisfying %the Clifford algebra
$\{\gamma^\mu,\gamma^\nu\}=2\eta^{\mu\nu}$
can be represented by the Pauli matrices
%$\gamma^0=\sigma_3=\sigma_z$ and $\gamma^1=-i\sigma_1=-i\sigma_x$.
$\gamma^0=\sigma_3$ and $\gamma^1=-i\sigma_1$.
Since in one spatial dimension there is no magnetic field
we can choose the gauge $qA_0=\Phi$ and $A_1=0$.
%Furthermore, we can choose the gauge $qA_0=\Phi$ and $A_1=0$
%(because in one spatial dimension there is no magnetic field ).
%
As a result, the Dirac equation simplifies to
\begin{equation}
i\d_t\Psi(t,x) = (-i \sigma_2 \d_x  + M \sigma_3 + \Phi) \Psi(t,x).
\end{equation}
In one spatial dimension, there is also no spin, hence the wave function
has only two components $\Psi=(\Psi^1,\Psi^2)$.
The Hamiltonian for the classical Dirac field reads then
\begin{equation}
H = \int dx\,{\Psi}\s (-i \sigma_2 \d_x  + M \sigma_3 + \Phi) \Psi
\,.
\end{equation}
As the next step, we discretize the space dimension and introduce a
regular grid (lattice) $x_n = n a$ with a positive grid (lattice)
constant $a$ and integers $n\in\mathbb Z$.
The discretization of the wave function $\Psi_n(t):=\sqrt{a}\,\Psi(t,x_n)$,
defined now at the grid points $x_n$, gives rise to a discretized derivative
$\sqrt{a}\,\d_x \Psi(t,x_n) \to [\Psi_{n+1}(t)-\Psi_{n-1}(t)]/(2a)$.
Finally, replacing the $x$-integral by a sum, we obtain
\begin{equation}
\label{continuum}
H \to \sum_n \Psi\s_n \left[-\frac{i\sigma_2}{2a} (\Psi_{n+1}-\Psi_{n-1})
+ M\sigma_3 \Psi_n + \Phi_n\Psi_n\right].
\end{equation}
In order to obtain the quantum many-body Hamiltonian, we quantize the
discretized Dirac field operators via the fermionic anti-commutation
relations
\begin{equation}
\{\h\Psi_n^\alpha(t),\h\Psi_m^\beta(t)\}=0
\;,\;\;%\quad
\{\h\Psi_n^\alpha(t),[\h\Psi_m^\beta(t)]\s\}=
\delta_{nm} \delta^{\alpha\beta}
\,.
\end{equation}
Using $\hat\Psi_n=(\hat a_n,\hat b_n)$, i.e., $\hat\Psi_n^1=\hat a_n$ and
$\hat\Psi_n^2=\hat b_n$, the discretized many-particle Hamiltonian reads
\bea
\label{H-lattice}
\h H
&=&
\frac1{2a} \sum_n
\left[ \h b^\dagger_{n+1} \h a_n - \h b^\dagger_n \h a_{n+1} + \hc \right]+
\nn
&&
+\sum_n \left[(\Phi_n+M)\h a\s_n \h a_n + (\Phi_n-M) \h b\s_n \h b_n \right]
\,.
\ea
The first term describes jumping between the neighboring grid points while the
remaining two terms can be treated as a combination of external potentials.
Due to the specific form of the jumping, the lattice splits into two
disconnected sub-lattices: (A) containing $\h a_{2n}$ and $\h b_{2n+1}$
and (B) containing $\h a_{2n+1}$ and $\h b_{2n}$ with integers $n$.
Since the two sub-lattices behave basically in the same way,
it is sufficient to consider only one of them, say A.
%
%
%
%Re-defining the local phases via
%$\h a_{2n} \to (-1)^n \h a_{2n}$ and 
%$\h b_{2n+1} \to (-1)^{n+1} \h b_{2n+1}$, 
%we obtain (for sub-lattice A)
%
%\bea
%\label{H-lattice-a-b}
%\h H
%&=&
%-\frac1{2a} \sum_n
%\left[ \h b\s_{2n+1} \h a_{2n} + \h b\s_{2n-1} \h a_{2n} + \hc \right]
%+
%\\
%&&
%+\sum_n
%\left[
%(\Phi_n+M) \h a\s_{2n} \h a_{2n} +(\Phi_n-M) \h b\s_{2n+1} \h b_{2n+1}
%\right]
%\,.
%\nonumber
%\ea
%
Identifying 
$ \h c_{2n} = (-1)^n\h a_{2n} $ and 
$ \h c_{2n+1} = (-1)^{n+1} \h b_{2n+1}$,
we obtain the form of the well-known Fermi-Hubbard Hamiltonian 
for a one-dimensional lattice
\begin{equation}
\label{Fermi-Hubbard0}
\h H = -J \sum_{n}
\left[\h c\s_{n+1} \h c_n + \h c\s_{n} \h c_{n+1}\right] +
\sum_n V_n \h c\s_n \h c_n
\,,
\end{equation}
with hopping rate $J=1/(2a)$ and on-site potentials 
$V_{2n}=\Phi_{2n}+M$ 
and 
$V_{2n+1}=\Phi_{2n+1}-M$. 
This Hamiltonian will be the starting point for the design of the 
optical lattice analogy.
But before we proceed, we note that the free part $\hat H_0$
of this Hamiltonian,
i.e., with $\Phi_n=0$, can be explicitly diagonalized.
Performing a discrete Fourier transform on the lattice
\begin{equation}
\label{a-b-Fourier}
\h a_p := \sum_n e^{-2inap}\, \h a_{2n}
\,,\;
\h b_p := \sum_n e^{-i(2n+1)ap}\, \h b_{2n+1}
\,,
\end{equation}
for $p\in(-\pi/2a,\pi/2a)$ and introducing new operators
\bea
\label{diag-a(p)}
  \begin{pmatrix}
    \hat A_p \\ \hat B_p
  \end{pmatrix}
=
\frac1{\sqrt{2E_p}}
\begin{pmatrix}
\sqrt{E_p+M} & \sqrt{E_p-M} \\
-\sqrt{E_p-M} & \sqrt{E_p+M}
\end{pmatrix}
\cdot
  \begin{pmatrix}
    \h a_p \\ \h b_p
  \end{pmatrix}
\,,
%\nn
%&=&
%U_p\cdot
%  \begin{pmatrix}
%    \h a_p \\ \h b_p
%  \end{pmatrix}
%\,,
\ea
we obtain the diagonalized form
\begin{equation}
\label{H0-a-b-sqrt}
\h H_0 = \int%\limits^{+\pi/(2a)}_{-\pi/(2a)}
dp\,
E_p
\left[\hat A_p^\dagger\hat A_p - \hat B_p^\dagger\hat B_p\right]
%\,.
\end{equation}
with the energy spectrum
\bea
\label{E_p}
E_p=\sqrt{M^2+\frac1{4a^2}\cos^2(ap)}
\,.
\ea
It approximates the relativistic energy-momentum relation at the edge 
of the Brillouin zone, for $p\approx \pm\pi/(2a)$.
The spectrum of $\h H_0$ consists thus of two symmetric intervals
separated by a gap of $2M$.
In order to obtain a positive Hamiltonian, we perform the usual
re-definition $\hat B_p^\dagger\leftrightarrow\hat B_p$ which
corresponds to changing the vacuum state by filling all $\hat B_p$
states by a fermion.
This is completely analogous to the \textit{Dirac sea} picture in quantum
electrodynamics.
In terms of this analogy, $\hat A_p^\dagger$ or $\hat A_p$ create or
annihilate an electron whereas $\hat B_p$ or $\hat B_p^\dagger$
create or annihilate a hole in the \textit{Dirac sea} -- 
which is then a positron.
An additional potential $\Phi_n$, if sufficiently localized in space,
will not modify this spectrum but may introduce
isolated eigenvalues with eigenstates corresponding to bound states
localized in space.

\paragraph{Experimental set-up}
The Fermi-Hubbard Hamiltonian (\ref{Fermi-Hubbard0}) can be realized
with ultra-cold fermionic atoms in a one-dimensional optical lattice
with the potential
\bea
 W(x) = W_0 \sin^2(2kx) + \Delta W \sin^2(kx)
\,,
\ea
where $k=\pi/(2a)$, by taking $W_0\gg\Delta W$.
Similar settings have already been obtained experimentally
\cite{Weitz-DoublePeriodicOptPot}.

%%%%%%%%%%%%%%%%%%%%%%%%%%%%%%%%%%%%%%%%%%%%%%%%%%%%%%%%%%%%%%%%%%%%%%%%%%%%%%%
\begin{figure}[h]
\includegraphics[width=\linewidth]{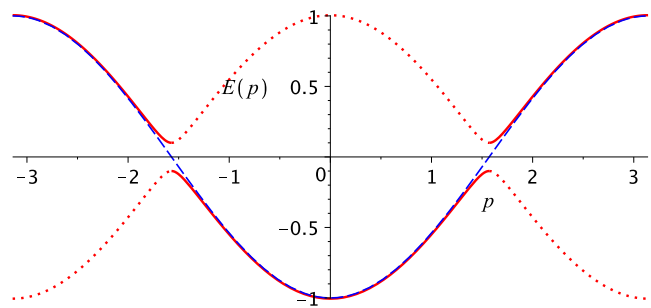}
\caption{Sketch of the dispersion relation (in units of $J$ and $1/a$)
for $\Delta W=0$ (dashed blue curve) and small $\Delta W>0$
(solid and dotted red curves).}
\label{fig:band-split}
\end{figure}
%%%%%%%%%%%%%%%%%%%%%%%%%%%%%%%%%%%%%%%%%%%%%%%%%%%%%%%%%%%%%%%%%%%%%%%%%%%%%%%

Although discretization of $W(x)$ at $x_n=na$ gives directly $V_n$
of \eqref{Fermi-Hubbard0}, the small perturbation $\Delta W$ doubles
the original periodicity of the potential $W(x)$ and thus some mathematical
caution in treating $\Delta W$ perturbatively is required.
By a version of WKB approximation for periodic potentials \cite{Balazs-WKB},
we re-derive the Hubbard model and the energy band structure from first
principles, using two sets of Wannier functions localized in the ``upper''
and the ``lower'' minima of the potential $W(x)$, respectively.
An interesting universal phenomenon occurs, see Fig.~\ref{fig:band-split}.
For $\Delta W=0$, the lowest band is described, in the
\textit{nearest-neighbor approximation} used here, by $E_p=-J\cos(ap)$
(the dashed blue curve in Fig.~\ref{fig:band-split}).
Switching on a small $\Delta W>0$, which immediately doubles the period 
of the potential, forces the energy dispersion relation to halve its 
period (the Brillouin zone shrinks by a factor of two) while keeping similar 
functional dependence on $p$ when $\Delta W \ll W_0$.
It leads to splitting of this band into two sub-bands
(red solid and dotted curves in Fig.~\ref{fig:band-split})
which are approximately described by
\begin{equation}
  E_{\pm}(p) = \pm\sqrt{M^2+J^2\cos^2(ap)}
\nonumber
\,.
\end{equation}
This nearest-neighbor approximation reproduces the relation~\eqref{E_p} 
for $J=1/(2a)$ and holds uniformly for all values of the quasi-momentum 
$p$ as long as $\Delta W$ and $J$ are small
%where $T$ is the transmission probability through the potential barrier 
\footnote{N.~Szpak and R.~Sch{\"u}tzhold, in preparation.}. 
In the vicinity of the minimum of the upper band and the maximum
of the lower band, separated by a gap $2M\approx\Delta W$, 
it reproduces the relativistic energy-momentum relation.
The corresponding Hamiltonian has the same form as the one for the 
discretized Dirac equation \eqref{H0-a-b-sqrt}, thus completing the analogy.

Using again the WKB approximation, we find that the hopping rate $J$ is mainly
determined by the potential strength $W_0$, the laser wave-number $k$
and the mass $m_{\rm atom}$ of the atoms (not to be confused with the
effective mass $M$ of the Dirac particles to be simulated)
\begin{align}
J \approx
\frac{4}{\pi}\,\sqrt{W_0 E_R}\,
\exp\left\{-\frac{\pi}{4}\sqrt{\frac{W_0}{E_R}}\right\},
\end{align}
where $E_R=k^2/(2m_{\rm atom})=\pi^2/(8m_{\rm atom}a^2)$ is the recoil energy.
The correction $\Delta W$ generates the mass gap $M \approx {\Delta W}/{2}$.
The analogue of the $e^+e^-$ pair creation can be simulated
if the involved scales obey the hierarchy
\bea
\omega_{\rm osc} \gg J \gg M \gg T
\,.
\ea
First, the local oscillator frequency $\omega_{\rm osc}$ in the
potential minima must be larger than $J$ to ensure the applicability
of the single-band Fermi-Hubbard Hamiltonian~(\ref{Fermi-Hubbard0}).
Second, $J \gg M$ is required for the continuum limit, i.e.,
that the discretized expression~(\ref{continuum}) provides a good
approximation.
Similarly, the change $\Delta\Phi_n=\Phi_{n+1}-\Phi_n$ of the
analogue of the electrostatic potential $\Phi_n$ from one site
to the next should be much smaller than $M$.
Over many sites, however, this change can well exceed the mass
gap $2M$, which is basically one of the conditions for the
Schwinger effect to occur.
Finally, the effective temperature $T$ should be well below the
mass gap $2M$ in order to avoid thermal excitations.

Let us insert some example parameters.
For ${}^6$Li atoms in an optical lattice made of light with a wavelength
of 500~nm, the recoil energy $E_R$ would be around $7\,\mu\rm K$.
If we choose the potential strength as $W_0=10\,\mu\rm K$, the
hopping rate $J$ would be around $5\,\mu\rm K$ which is still
sufficiently below the local oscillator frequency $\omega_{\rm osc}$
of around $34\,\mu\rm K$.
With $\Delta W=1\,\mu\rm K$ we would get a mass $M$ of 500~nK
and the effective temperature should be below that value.

\paragraph{Bose-Fermi mapping}
Since it is typically easier to cool down bosonic than fermionic atoms,
let us discuss an alternative realization based on bosons in an optical
lattice.
To this end, we start with the Bose-Hubbard Hamiltonian which has the
same form as the Fermi-Hubbard Hamiltonian \eqref{Fermi-Hubbard0}
after replacing the fermionic $\hat c_n$ by bosonic $\hat d_n$ operators,
but with an additional on-site repulsion term
$U\,(\hat d^\dagger_n\hat d_n-1)\,\hat d^\dagger_n\hat d_n$.
For large $U\gg J$ (which can be controlled by an external magnetic field
via a Feshbach resonance, for example), we obtain the bosonic analogue
of ``Pauli blocking'', i.e., at most one particle can occupy each site.
Neglecting all states with double or higher occupancy, we can map these
bosons exactly onto fermions in one spatial dimension via
\begin{equation}
\h c_n = \h d_n \prod_{m<n} \exp\left(-i\pi\, \h d\s_m \h d_m\right)
%\h d_n = \h c_n \prod_{m<n} \exp\left(-i\pi \h c\s_m \h c_m\right)
%\h d_n = \exp\left(-i\pi\sum_{m<n} \h c\s_m \h c_m\right) \h c_n
\,.
\end{equation}
As a result, we obtain the same physics as described by the
Fermi-Hubbard Hamiltonian \eqref{Fermi-Hubbard0}.

%%%%%%%%%%%%%%%%%%%%%%%%%%%%%%%%%%%%%%%%%%%%%%%%%%%%%%%%%%%%%%%%%%%%%%%%%%%%%%%
\begin{figure}[h]
\includegraphics[width=1\linewidth,height=0.1\textheight]{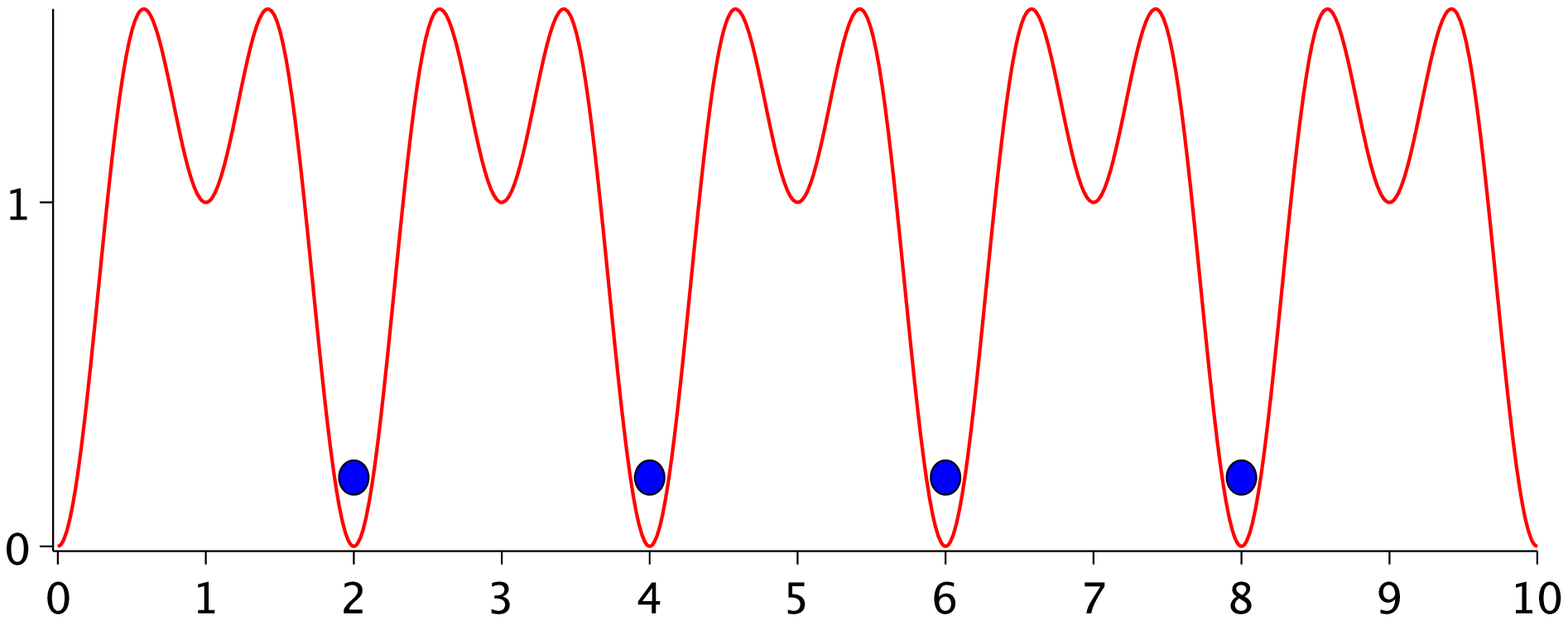}
\includegraphics[width=1\linewidth,height=0.1\textheight]{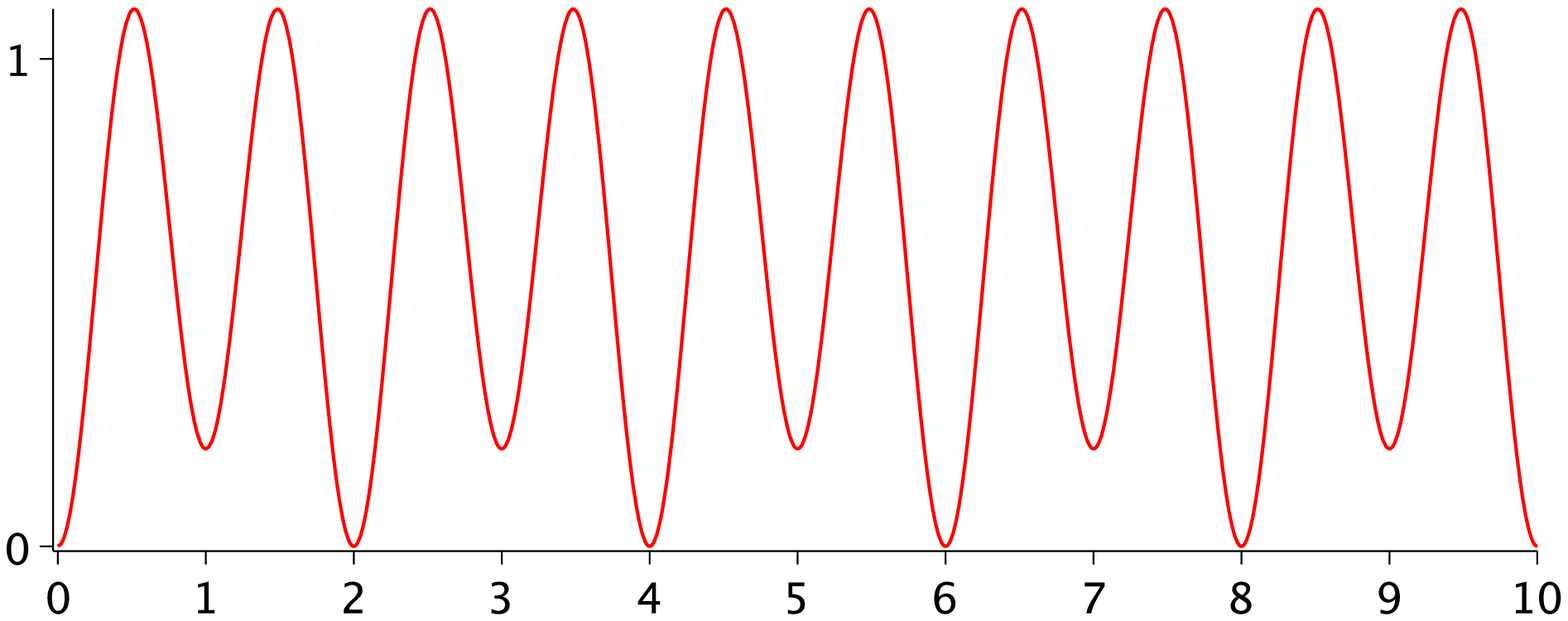}
\includegraphics[width=1\linewidth,height=0.15\textheight]{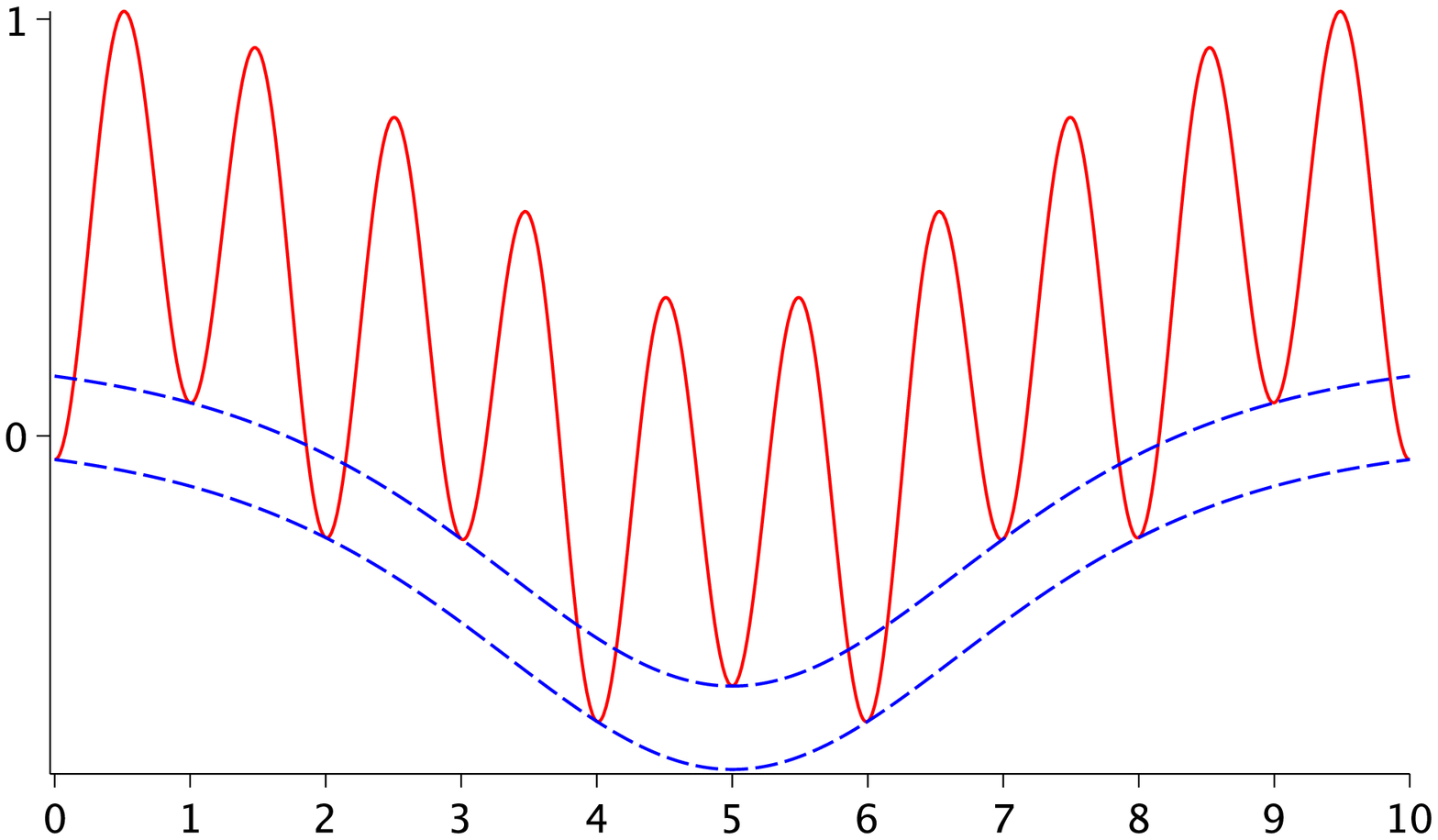}
\includegraphics[width=1\linewidth,height=0.1\textheight]{dppot-low}
\includegraphics[width=1\linewidth,height=0.1\textheight]{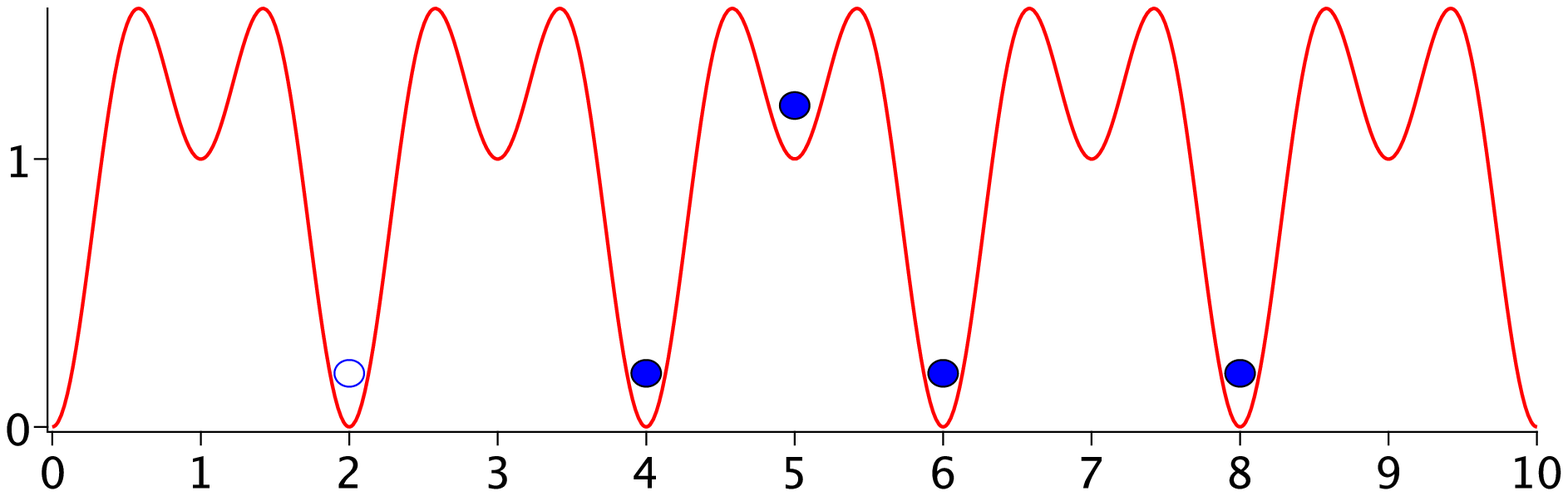}
\caption{Sketch (not to scale) of the temporal stages of the simulation
(from top to bottom).
The solid red curves represent the optical lattice potentials $W(x)$
as a function of position $x$ and the dashed blue curves correspond
to the effective electric potential $\Phi$.
The blue solid dots are particles and the empty (blue) circle is a hole.}
\label{Fig:stages}
\end{figure}
%%%%%%%%%%%%%%%%%%%%%%%%%%%%%%%%%%%%%%%%%%%%%%%%%%%%%%%%%%%%%%%%%%%%%%%%%%%%%%%

\paragraph{Simulation procedure}
The above established analogy between the (discretized) quantum many-particle
Hamiltonian of electrons and positrons in an electric field, on the one hand,
and the (Bose or Fermi) Hubbard model describing ultra-cold atoms in optical
lattices, on the other hand, enables laboratory simulations of relativistic
phenomena of strong-field QED.
Probably the most prominent of them is the spontaneous pair creation in
strong electric fields known also as the Schwinger effect
\cite{Schwinger, Sauter, Heisenberg+Euler} which has to date not been
confirmed experimentally.
The original Schwinger effect \cite{Schwinger} deals with a constant
electric field $E$ and would correspond to a static tilted optical lattice
with $\Phi(x) = E x$.
However, in view of the planned experiments \cite{ELI}, an electric field 
which is localized in space and time is more realistic.
In order to clearly distinguish non-perturbative spontaneous pair creation 
(via tunneling) from other perturbative effects such as dynamical pair 
creation, the electric field should be slowly varying (compared to the
frequency scale of the gap $2M$).

As a result, we envisage an experimental realization sketched in
Fig.~\ref{Fig:stages}.
In order to prepare the initial state, we start with $\Delta W\gg W_0$
where the two bands are separated by a large gap.
The \textit{Dirac sea} then corresponds to the state where all lower minima 
are filled with atoms while all upper minima are empty
(half filling of the lattice, see first picture in Fig.~\ref{Fig:stages}).
If we then decrease $\Delta W$ adiabatically until $\Delta W\ll W_0$
and thus $J\gg M$, the atoms become delocalized --
but still the lower  band is filled while the upper band remains empty
(second picture in Fig.~\ref{Fig:stages}).
As the next stage, we slowly switch on an additional 
potential $\Phi$ to facilitate tunneling from the lower band to
the upper band -- the analogue of the Schwinger effect
(third picture in Fig.~\ref{Fig:stages}).
Finally, after slowly switching off the potential $\Phi$ again
(fourth picture in Fig.~\ref{Fig:stages}), we increase $\Delta W$
adiabatically until $\Delta W\gg W_0$.
By this energetic separation, a particle-hole pair created via tunneling 
is transformed to an
atom in one of the upper minima and, consequently, a missing atom
(i.e., hole) in one of the lower minima 
(fifth picture in Fig.~\ref{Fig:stages}).
This could be detected by site-resolved imaging
\cite{Bloch-SingleAtomResolvedFluorescence}, for example.

Note that the creation of a particle-hole pair separated by $\Delta n$
lattice sites requires the simultaneous tunnelling of $\Delta n$ 
particles since two particles are not allowed to occupy the same site. 
This again emphasizes the many-particle character of our proposal which 
goes beyond the simulation of the classical Dirac equation.
Apart from investigating the pair creation probability for
space-time dependent electric fields $E(t,x)$, this quantum
simulator for the Schwinger effect could provide some insight
into the impact of interactions.
For example, including dipolar interactions of the atoms, we may
even switch between attractive and repulsive interactions by
aligning the atoms spins parallel or perpendicular to the lattice.

\bigskip

R.~S.\ acknowledges stimulating discussions during the 
Benasque Workshop on {\em Quantum Simulations} in 2011
and financial support from the DFG (SFB-TR12).

%%%%%%%%%%%%%%%%%%%%%%%%%%%%%%%%%%%%%%%%%%%%%%%%%%%%%%%%%%%%%%%%%%%%%%%%%%%%%%%
% \bibliography{basename of .bib file}
%\bibliography{lattices}
%\bibliographystyle{unsrt}
%\bibliographystyle{apsrev}

\end{document}